\documentclass[]{spie}  

\usepackage{amsmath,amsfonts,amssymb}
\usepackage{graphicx}
\usepackage[colorlinks=true, allcolors=blue]{hyperref}

\usepackage{xcolor}

\title{Demonstration of simultaneous PIAA- coronagraphy and wavefront sensing using a single metasurface-based focal-plane optic}

\author[a]{Dhwanil Patel}
\author[a, c]{Sebastiaan Y. Haffert}
\author[b]{Skyler Palatnick}
\author[a]{Adam Taras}
\author[b]{Maxwell A. Millar-Blanchaer}
\author[a]{Matthijs Mars}
\author[a]{Elena Tonucci}
\author[c]{Jared R. Males}
\author[c]{Laird M. Close}
\author[e]{Joshua Liberman}
\author[c]{Warren B. Foster}
\author[c]{Kyle Van Gorkom}
\author[c,e,f,g]{Olivier Guyon}
\author[h]{Alexander D. Hedglen}
\author[c]{Parker T. Johnson}
\author[e]{Maggie Y. Kautz}
\author[e]{Jay K. Kueny}
\author[c]{Jialin Li}
\author[d]{Joseph D. Long}
\author[e]{Jennifer Lumbres}
\author[e]{Eden A. McEwen}
\author[i]{Avalon McLeod}
\author[j]{Lauren Schatz}
\author[e]{Katie Twitchell}
\author[k]{Robert J. Harris}
\author[k]{Viktoria Kutnohorsky}

\affil[a]{Leiden Observatory, Leiden University, PO Box 9513, 2300 RA Leiden, The Netherlands}
\affil[b]{Department of Physics, University of California, Santa Barbara, CA 93106, USA}
\affil[c]{Steward Observatory, University of Arizona, 933 North Cherry Avenue, Tucson, AZ 85721, USA}
\affil[d]{Center for Computational Astrophysics, Flatiron Institute, 162 5th Avenue, New York, NY 10010, USA}
\affil[e]{Wyant College of Optical Sciences, University of Arizona, 1630 E University Blvd., Tucson, AZ 85721, USA}
\affil[f]{Subaru Telescope, National Astronomical Observatory of Japan, 650 N. A'ohoku Place, Hilo, HI 96720, USA}
\affil[g]{Astrobiology Center, National Institutes of Natural Sciences, Mitaka, Tokyo, Japan}
\affil[h]{Northrop Grumman Corporation, Rolling Meadows, IL, USA}
\affil[i]{Draper Laboratory, Cambridge, MA, USA}
\affil[j]{Starfire Optical Range, Kirtland Air Force Base, Albuquerque, NM, USA}
\affil[k]{Department of Physics, Durham University, Durham, United Kingdom}

\authorinfo{Further author information: (Send correspondence to D.P.)\\ D.P.: E-mail: dpatel@strw.leidenuniv.nl }

\begin{document} 
\maketitle

\begin{abstract}
Controlling the residual wavefront aberrations downstream of an extreme adaptive optics (ExAO) system is one of the main challenges in high-contrast imaging. The residual aberrations produce quasi-static speckles, mainly due to the difference between the paths taken by the wavefront-sensing light and the science light after they split in an ExAO system. Highly sensitive wavefront sensors, such as Zernike wavefront sensors (ZWFSs), are used as second-stage correctors to mitigate these non-common path aberrations. They are implemented as focal-plane optics in the science path. High-performing coronagraphs, such as complex mask coronagraphs (CMCs), are also implemented in the focal plane. They also provide a small ($\lesssim 1.5\,\lambda/D$) inner working angle compared to other coronagraphs. Both have been shown to perform better when used with lossless apodization such as phase-induced amplitude apodization (PIAA) optics. Hence, it is natural to combine the two functionalities into a single focal-plane optic. Metasurfaces can be designed to have chromatic responses, and hence they can be used to realize such a focal-plane optic with different functionalities in different wavelength bands. We demonstrate the viability of such an optic by manufacturing a hybrid metasurface that is designed to function as a CMC and a ZWFS in two intermediate-band filters in the H band, each with a fractional bandwidth of approximately 1\%. Specifically, we show the measured optical responses at different wavelengths in the band, demonstrating different phases with $\sim \pi/2$ at shorter wavelengths and $\pi$ at longer wavelengths between $1500$ and $1700\,\text{nm}$. This would allow for wavefront sensing at the shorter wavelength of $\sim1500\,\text{nm}$ and coronagraphy at the longer wavelength of $\sim1700\,\text{nm}$. Additionally, we tested the mask on-sky with the MagAO-X instrument at the Magellan Clay 6.5 m telescope at Las Campanas Observatory, Chile. On-sky results show a contrast of $\sim 10^{-1}$ at a non-ideal wavelength of $\sim 1600\,\text{nm}$. This is comparable to the contrast curves from the simulated performance using the optical responses obtained from metrology around that wavelength. Finally, we evaluated the wavefront-sensing performance of the fabricated metasurface using the MagAO-X internal source at $1300\,\mathrm{nm}$ and compared it with simulations. The measured reconstruction error is consistent with simulations of an ideal Zernike wavefront sensor, confirming the wavefront-sensing functionality of the metasurface.

\end{abstract}


\section{Introduction}
\label{sect:intro}
Directly imaging and studying exoplanets requires very high sensitivity and angular resolution. The next generation of extremely large telescopes will be able to provide these capabilities. However, ground-based telescopes are affected by atmospheric turbulence, which reduces both sensitivity and resolution. To mitigate this effect, high-contrast imaging (HCI) instruments use extreme adaptive optics (ExAO) systems \cite{guyon2018extreme}. Current HCI instruments implemented on 6.5–10 meter class telescopes, such as SPHERE and NaCo on the VLT \cite{fusco2016saxo}, ScExAO on the Subaru Telescope \cite{jovanovic2015subaru}, MagAO-X on the Magellan Clay Telescope \cite{males2018magao}, and GPI on the Gemini South telescope \cite{macintosh2014first}, combined with advanced post-processing techniques, have enabled the direct detection of planetary-mass companions \cite{lagrange2010giant, macintosh2015discovery, haffert2019two, currie2023direct, lucas2024visible}. However, reflected-light characterization of smaller and closer-in planets requires deeper contrast and smaller inner working angles.

\begin{figure}[hbt!]
    \centering
    \includegraphics[width=\linewidth]{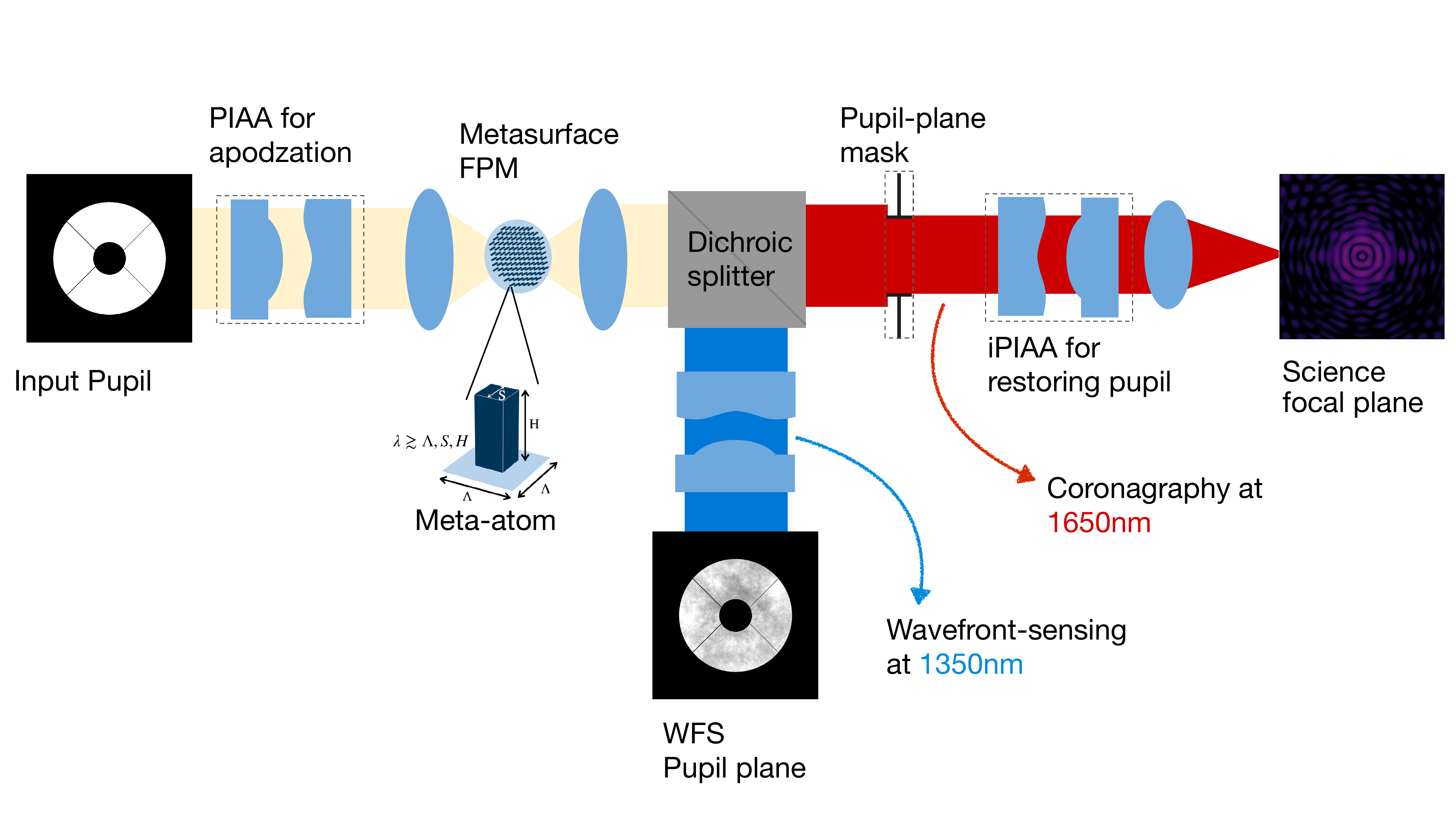}
    \caption{Schematic of the proposed architecture. The input pupil is first apodized by a set of PIAA lenses and focused onto a chromatic metasurface focal-plane mask (FPM). The metasurface imparts wavelength-dependent phase shifts to the PSF core, enabling wavefront sensing in one spectral band and coronagraphy in another. A dichroic beam splitter downstream of the FPM separates the light into wavefront-sensing and coronagraphic channels. In the coronagraphic channel, the beam first propagates through a Lyot stop to block starlight diffracted by the coronagraph. In both channels, the beam then passes through dedicated inverse-PIAA optics that restore the original pupil geometry. Finally, the coronagraphic beam is focused onto the science detector, while the wavefront-sensing beam is relayed to a pupil-plane detector for wavefront measurements.}
    \label{fig:optical_setup}
\end{figure}

One of the main limitations to achieving deeper contrast is the presence of non-common path aberrations (NCPAs) in ExAO systems. NCPAs arise due to differences between the optical paths taken by the wavefront-sensing light and the science light after they split inside the instrument. These aberrations are generated within the instrument and evolve slowly over time due to thermal, mechanical, environmental, and gravity-vector changes, resulting in quasi-static speckles in the science focal plane.
Several concepts have been proposed to sense and minimize NCPAs using wavefront sensing close to, or within, the science optical path \cite{pourcelot2022low, soummer2022high, ruane2023dual, wallace2024dual, chambouleyron2024coronagraph}. Among these, the Zernike wavefront sensor (ZWFS), based on the Zernike phase contrast technique \cite{zernike1935phase}, provides very high sensitivity \cite{n2013calibration}. It behaves as an interferometric wavefront sensor, where a phase-shifting mask modifies part of the point spread function (PSF) to generate a reference field that interferes with the remaining beam and produces an intensity signal proportional to phase aberrations. Since the phase mask is transmissive, the ZWFS maintains high photon efficiency \cite{chambouleyron2021variation}, enabling picometer-level sensitivity \cite{ruane2020wavefront, steeves2020picometer}.

At the same time, achieving high raw contrast also requires efficient suppression of on-axis starlight. Complex mask coronagraphs (CMCs) have demonstrated deep contrast at small inner working angles while maintaining high throughput within the phase-induced amplitude apodization (PIAA) framework \cite{guyon2006theoretical, guyon2010high, belikov2022laboratory}. Similar to a ZWFS, a CMC uses a transmissive focal-plane mask to phase-shift part of the PSF. However, the purpose of the phase shift differs: in a CMC, the shifted and unshifted fields interfere destructively to suppress starlight, whereas in a ZWFS, they interfere constructively to generate a wavefront-sensing signal. In their simplest forms, a CMC uses a uniform $\pi$ phase shift and a ZWFS uses a $\pi/2$ phase shift. 

Metasurfaces use arrays of subwavelength structures to engineer the optical response of light and can be designed to exhibit chromatic behaviour \cite{yu2014flat}. As a result, different optical functionalities can be realized in different wavelength bands using a single optical element. This enables coronagraphy and wavefront sensing to share a common focal-plane mask, as shown schematically in Figure \ref{fig:optical_setup}, thereby minimizing non-common optical paths between the two channels \cite{haffert2023integrated}.

In this work, we experimentally validate a hybrid metasurface that combines coronagraphy and wavefront sensing at a common focal plane. The metasurface is designed to operate as a Zernike wavefront sensor in one intermediate-band filter and as a complex-mask coronagraph in another intermediate-band filter, both within the H band. Section \ref{sec: maskDesign} describes the design of the metasurface-based focal-plane mask together with the manufacturing process and laboratory metrology used to characterize its optical response at different wavelengths. Section \ref{sec: onsky_results} and \ref{sec: lab_wfs} show the on-sky coronagraphic performance and in-lab wavefront-sensing performance, respectively. Finally, section \ref{sec: conclusion} discusses future directions. 

\section{Focal-plane mask}
\label{sec: maskDesign} 
\subsection{Metasurface design}
As a prototype, we designed a simple metasurface consisting of identical amorphous silicon (a-Si) meta-atoms patterned on a silica ($\mathrm{SiO_2}$) substrate. This produces a spatially uniform near-field phase response, allowing the device to function as a Zernike wavefront sensor ($\pi/2$ phase shift) at shorter wavelengths and as a Roddier \& Roddier phase-mask ($\pi$ phase shift) at longer wavelengths. A square nanopost for meta-atom geometry was chosen to make the metasurface polarization-insensitive. The meta-atom height was fixed at $H = 1.02\,\mu\mathrm{m}$, while the period ($\Lambda$) and lateral size ($S$) were treated as free parameters. These parameters were scanned using rigorous coupled-wave analysis (RCWA) over wavelengths spanning the J and H bands. From the parameter sweep, an optimal configuration was identified with $H = 1.020\,\mu\mathrm{m}$, $\Lambda = 0.802\,\mu\mathrm{m}$, and $S = 0.329\,\mu\mathrm{m}$. This design provides phase shifts of approximately $3.13\,\mathrm{rad}$ and $-1.61\,\mathrm{rad}$ at wavelengths of $1.57\,\mu\mathrm{m}$ and $1.42\,\mu\mathrm{m}$, respectively, while maintaining a transmission of $\gtrsim 90\%$. Therefore, the designed metasurface provides chromatic phase shifts close to $\pi$ and $\pi/2$ at the two target wavelengths with high transmission.

\begin{figure}[hbt!]
    \centering
    \includegraphics[width=\linewidth]{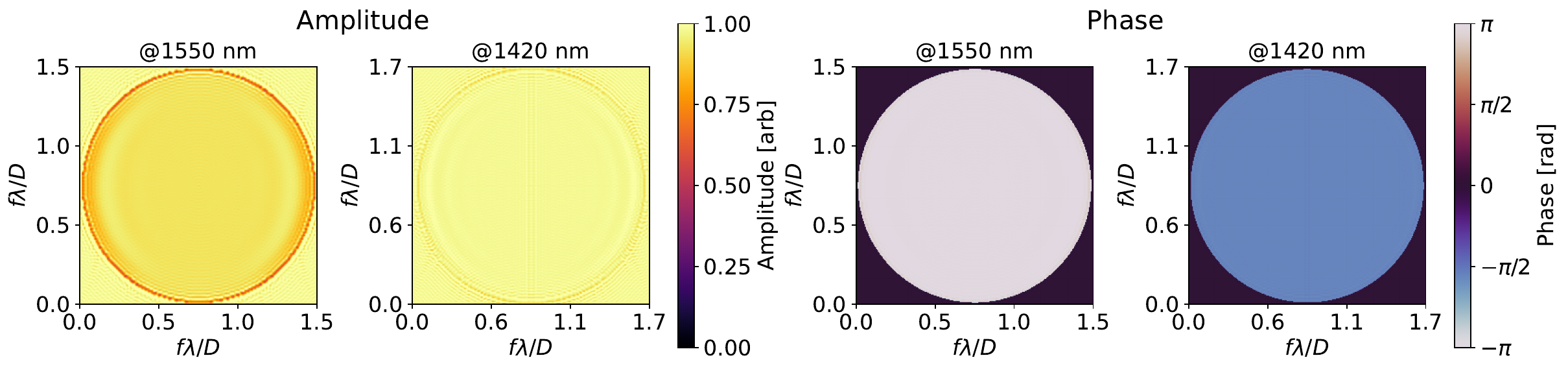}
    \caption{Near-field optical response of the metasurface obtained from FDTD simulations. The amplitude (left two panels) and phase (right two panels) responses of the metasurface at $1550\,\mathrm{nm}$ and $1420\,\mathrm{nm}$. The metasurface has a physical diameter of approximately $160\,\mu\mathrm{m}$, corresponding to $1.50\,\lambda/D$ and $1.69\,\lambda/D$ at the two wavelengths, respectively. At $1550\,\mathrm{nm}$ the metasurface provides an approximately uniform $\pi$ phase shift, while at $1420\,\mathrm{nm}$ it provides an approximately uniform $\pi/2$ phase shift. The amplitude response remains close to unity across most of the mask, with localized variations near the mask boundary.}
    \label{fig:fdtd_mask}
\end{figure}


A full metasurface consisting of the optimized meta-atom geometry was then simulated using the finite-difference time-domain (FDTD) method to calculate its near-field optical response. Figure \ref{fig:fdtd_mask} shows the simulated near-field response at $1.55\,\mu\mathrm{m}$ and $1.42\,\mu\mathrm{m}$ in the top and bottom rows, respectively. The phase maps are spatially uniform at both wavelengths, as expected. 
Some artifacts can be observed in the transmission amplitude maps, particularly near the mask boundaries. These arise because the periodic assumption used in the metasurface design no longer holds at the edges of a finite structure. At the boundary, the meta-atom array transitions abruptly to an unpatterned region, resulting in localized amplitude distortions along the edge of the masks. Additional high-frequency oscillations are also visible in some regions of the amplitude maps. These are likely due to resonant effects inherent in subwavelength dielectric structures \cite{yu2014flat} and should be considered when evaluating the performance of the metasurface. 

\subsection{Manufacturing and Characterization}
\label{sec: manufac} 
\begin{figure}[hbt!]
    \centering
    \includegraphics[width=\linewidth]{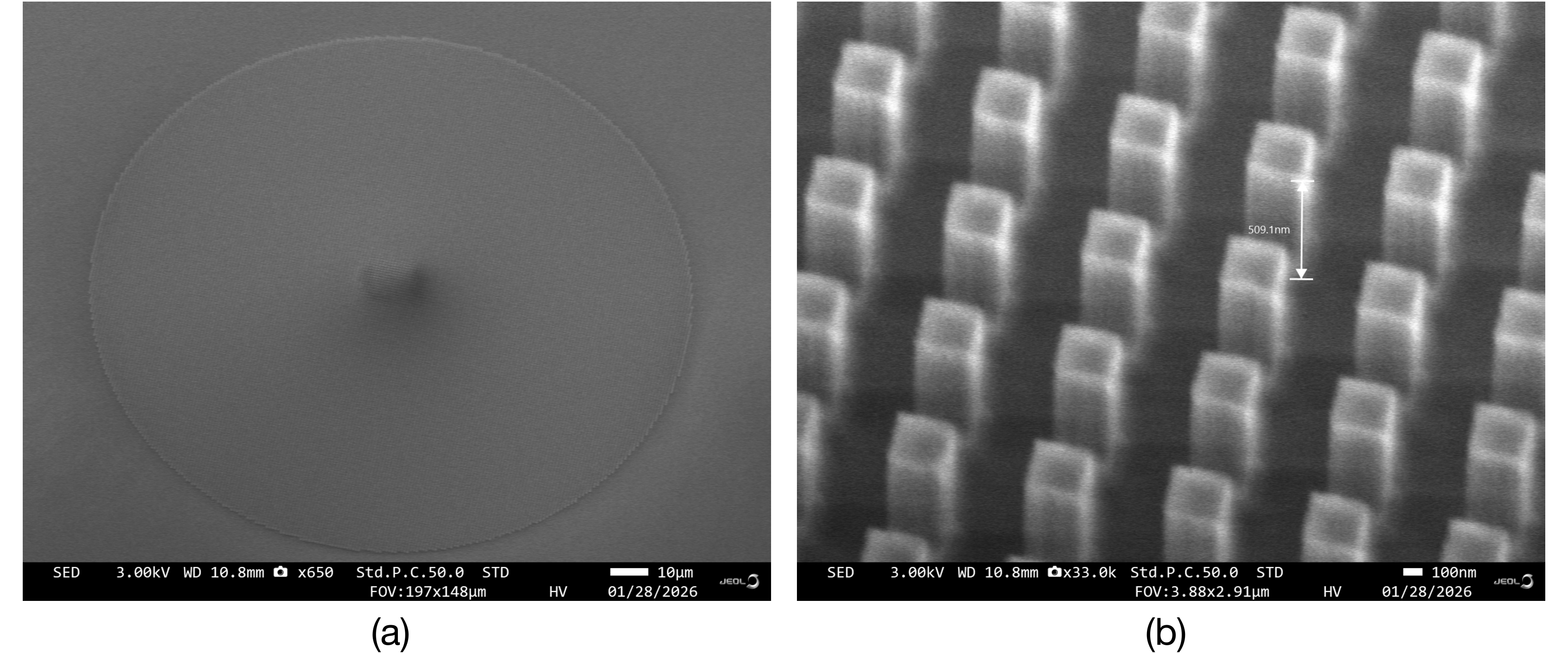}
    \caption{SEM images of the fabricated metasurface. (a) Overview of the metasurface phase mask. The dark feature in the center region is an SEM charging artifact rather than a fabrication defect. (b) Zoomed-in view of the repeating meta-atoms.}
    \label{fig:sem_mask}
\end{figure}

The optimized meta-atom geometry was subsequently fabricated as a proof-of-concept metasurface with a physical diameter of $\sim160\,\mu\mathrm{m}$, corresponding to $1.5\,\lambda/D$ at $1.55\,\mu\mathrm{m}$. The device was fabricated using electron-beam lithography at the UCSB Nanofabrication Facility. Figure~\ref{fig:sem_mask} shows scanning electron microscope images of the fabricated metasurface. 

To characterize the fabricated metasurface, its transmission and phase response were measured at multiple wavelengths using a digital holographic microscope (DHM) and a broadband infrared laser source. The measurement setup and reconstruction procedure were previously described in detail \cite{palatnick2024achromatizing, wallace2026short}. In this approach, light from a single-mode source is split into a reference beam and a sample beam that propagates through the metasurface. The resulting interference pattern is recorded and numerically reconstructed to recover the complex transmitted field, yielding both amplitude and phase information.

\begin{figure}[hbt!]
    \centering
    \includegraphics[width=\linewidth]{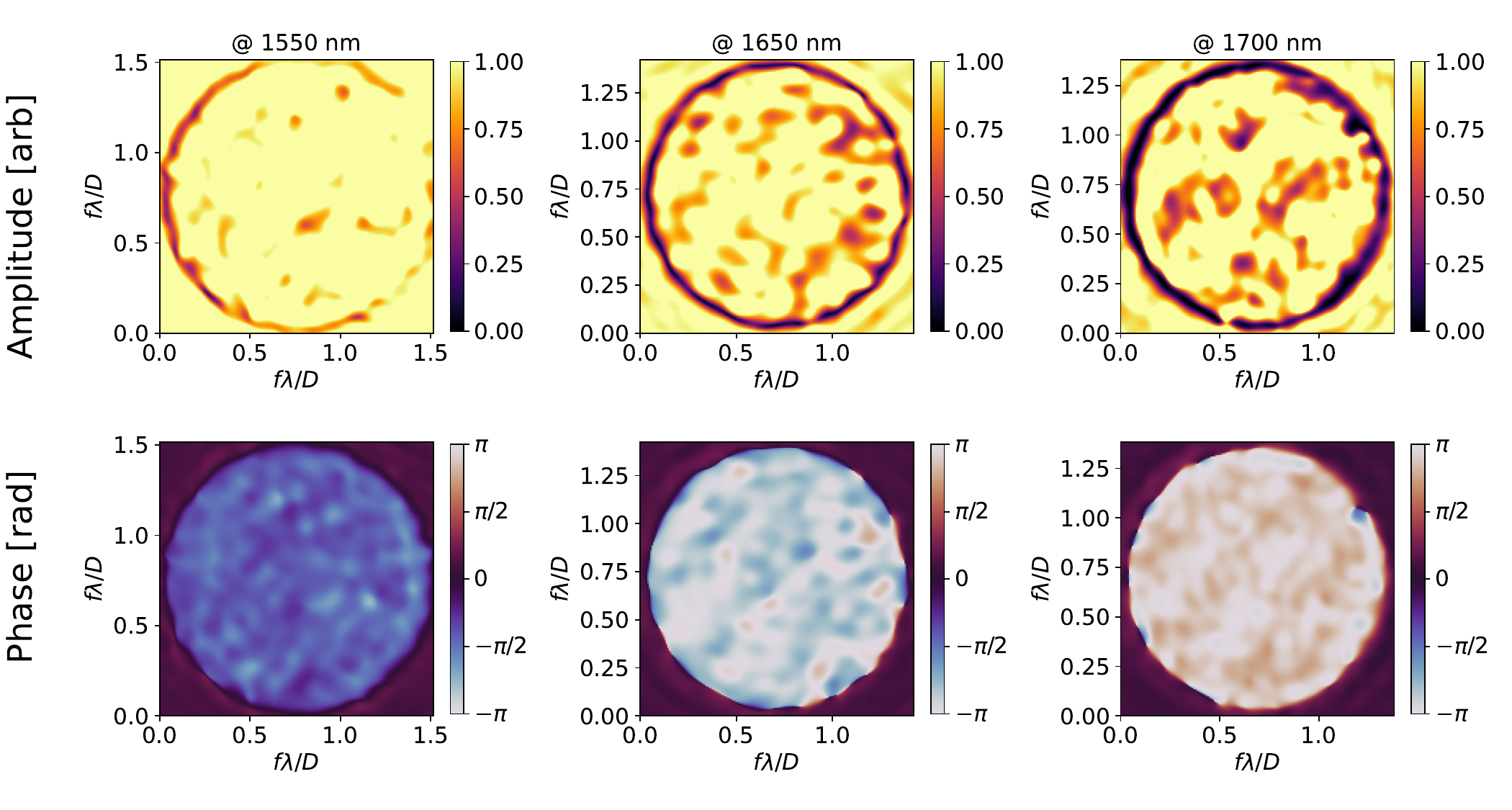}
    \caption{Near-field measurements of the metasurface. Amplitude (top-row) and phase (bottom-row) from the metasurface measured at $1550\,\mathrm{nm}$, $1650\,\mathrm{nm}$ and $1700\,\mathrm{nm}$ in the first, second and third column, respectively.}
    \label{fig:lab_mask_transmission_phase}
\end{figure}

Figure \ref{fig:lab_mask_transmission_phase} shows the measured phase and transmission of the fabricated metasurface at $1550\,\mathrm{nm}$, $1650\,\mathrm{nm}$, and $1700\,\mathrm{nm}$. The diameter of the central phase-shifting region is approximately $1.5\,\lambda/D$ at $1550\,\mathrm{nm}$, consistent with the design. The transmission maps show reduced transmission near the boundary between the metasurface region and the surrounding substrate, as expected from the abrupt transition between areas containing meta-atoms and areas without meta-atoms. Additional localized regions of reduced transmission are also present. However, these features occur on spatial scales much larger than the individual meta-atoms and appear at different locations for different wavelengths, suggesting they are unlikely to originate from the metasurface design or fabrication process. Instead, they may be associated with systematic effects in the DHM measurement. 

The measured phase response confirms that the metasurface imparts a strong wavelength-dependent phase shift. The device was designed to provide a relative phase shift of approximately $\pi$ at $1550\,\mathrm{nm}$ and $\pi/2$ at $1420\,\mathrm{nm}$. While measurements at shorter wavelengths were not available, the reconstructed phase maps indicate a phase shift closer to $\pi/2$ at $1550\,\mathrm{nm}$ and approaching $\pi$ at $1700\,\mathrm{nm}$. The origin of this apparent spectral shift is yet to be studied. Nevertheless, the measurements demonstrate that the metasurface produces the intended chromatic phase-shifting behavior with high transmission. 


\section{On-sky coronagraphic performance}
\label{sec: onsky_results}
The fabricated mask was tested on-sky using the MagAO-X instrument mounted on the 6.5-m Magellan Clay Telescope at Las Campanas Observatory. Observations targeted the p Velorum triple-star system. The primary component of the system is itself a spectroscopic binary (p Vel Aa and p Vel Ab), while the tertiary component (p Vel B) appears as a nearby companion in the science images \cite{evans1969rediscussion_pvel}. Throughout the remainder of this work, these sources are referred to as the primary and companion, respectively. The observations were obtained during the 2026A observing run on 28 March 2026, between 03:25 and 04:40 UTC (median seeing condition $\sim 0.62"$).

Reference non-coronagraphic images were first obtained with the metasurface FPM removed from the beam, providing measurements of the direct point spread function (PSF). This reference dataset consisted of 132 frames recorded with the infrared camera at a frame rate of 2 Hz. Coronagraphic observations were then performed with the metasurface FPM aligned in the science beam. The final coronagraphic dataset used for the analysis consisted of 74 frames recorded with the same camera at the same frame rate. The images were taken at $1600\,\mathrm{nm}$ at 1\% bandwidth, and hence the FPM would be $\sim 1.45\,\lambda/D$ \footnote{From now on, focal-plane separations are expressed in units of \(\lambda/D\), corresponding to a physical scale of \(f\lambda/D\)}. The dark-corrected median non-coronagraphic and coronagraphic images, computed from the full set of recorded frames, are shown in the top row (a) of Figure~\ref{fig:reduction_steps}.

\subsection{Data reduction}
\begin{figure}[hbt!]
    \centering
    \includegraphics[width=\linewidth]{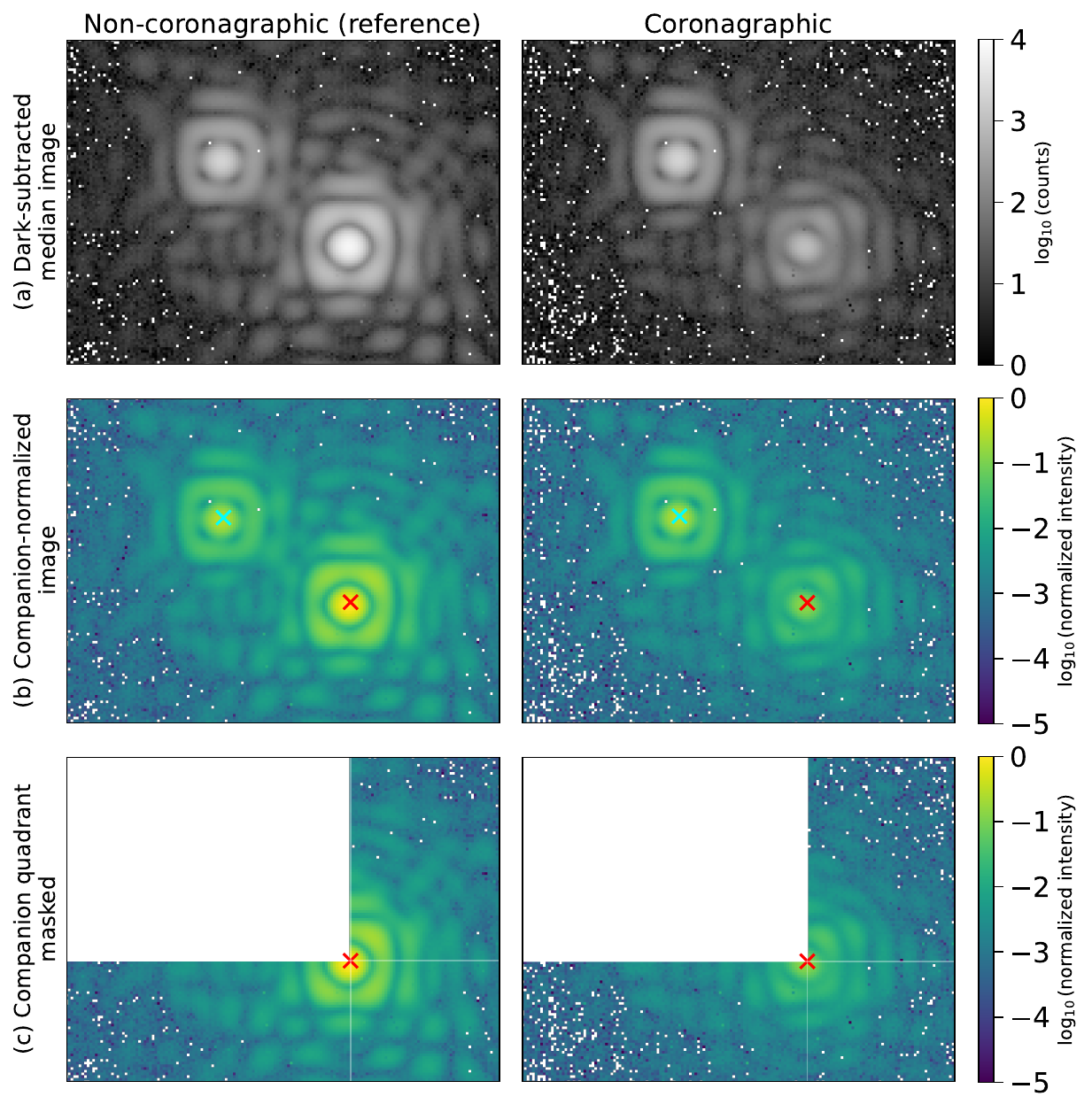}
    \caption{Data reduction steps for the on-sky observations. The left and right columns show the non-coronagraphic (reference) and coronagraphic observations, respectively. Row (a) shows the median-combined images after dark subtraction. Row (b) shows the corresponding normalized intensity images, where detector counts have been converted to normalized intensity using the companion source as a photometric reference. Row (c) shows the images after masking the quadrant containing the companion source.}
    \label{fig:reduction_steps}
\end{figure}

Photometric normalization was performed using the companion visible in the field of view. The companion-to-primary peak intensity ratio was first measured from the non-coronagraphic reference image,

\begin{equation}
R =
\frac{P_{\rm comp}^{\rm ref}}
     {P_{\rm prim}^{\rm ref}},
\label{eq:companion_ratio}
\end{equation}

where \(P_{\rm comp}^{\rm ref}\) and \(P_{\rm prim}^{\rm ref}\) are the measured companion and primary peak intensities, respectively. Assuming this ratio remains constant throughout the observations, the expected primary peak intensity in the coronagraphic image $\left(P_{\rm prim}^{\rm exp, cor}\right)$ is

\begin{equation}
P_{\rm prim}^{\rm exp, cor}
=
\frac{P_{\rm comp}^{\rm cor}}
     {R}
=
P_{\rm comp}^{\rm cor}
\frac{P_{\rm prim}^{\rm ref}}
     {P_{\rm comp}^{\rm ref}},
\label{eq:primary_estimate}
\end{equation}

where \(P_{\rm comp}^{\rm cor}\) is the measured companion peak intensity in the coronagraphic image. The normalized image is then given by

\begin{equation}
I_{\rm norm}(x,y)
=
\frac{I_{\rm med}(x,y)}
     {P_{\rm prim}^{\rm exp, cor}}.
\label{eq:normalization}
\end{equation}
The resulting normalized reference and coronagraphic images are shown in row (b) of Figure~\ref{fig:reduction_steps}.
To prevent contamination from the companion source, the quadrant of the image containing the companion was masked before further analysis. The masked images are shown in row (c) of Figure~\ref{fig:reduction_steps}.

\subsection{Contrast curves}
\begin{figure}[hbt!]
    \centering
    \includegraphics[width=\linewidth]{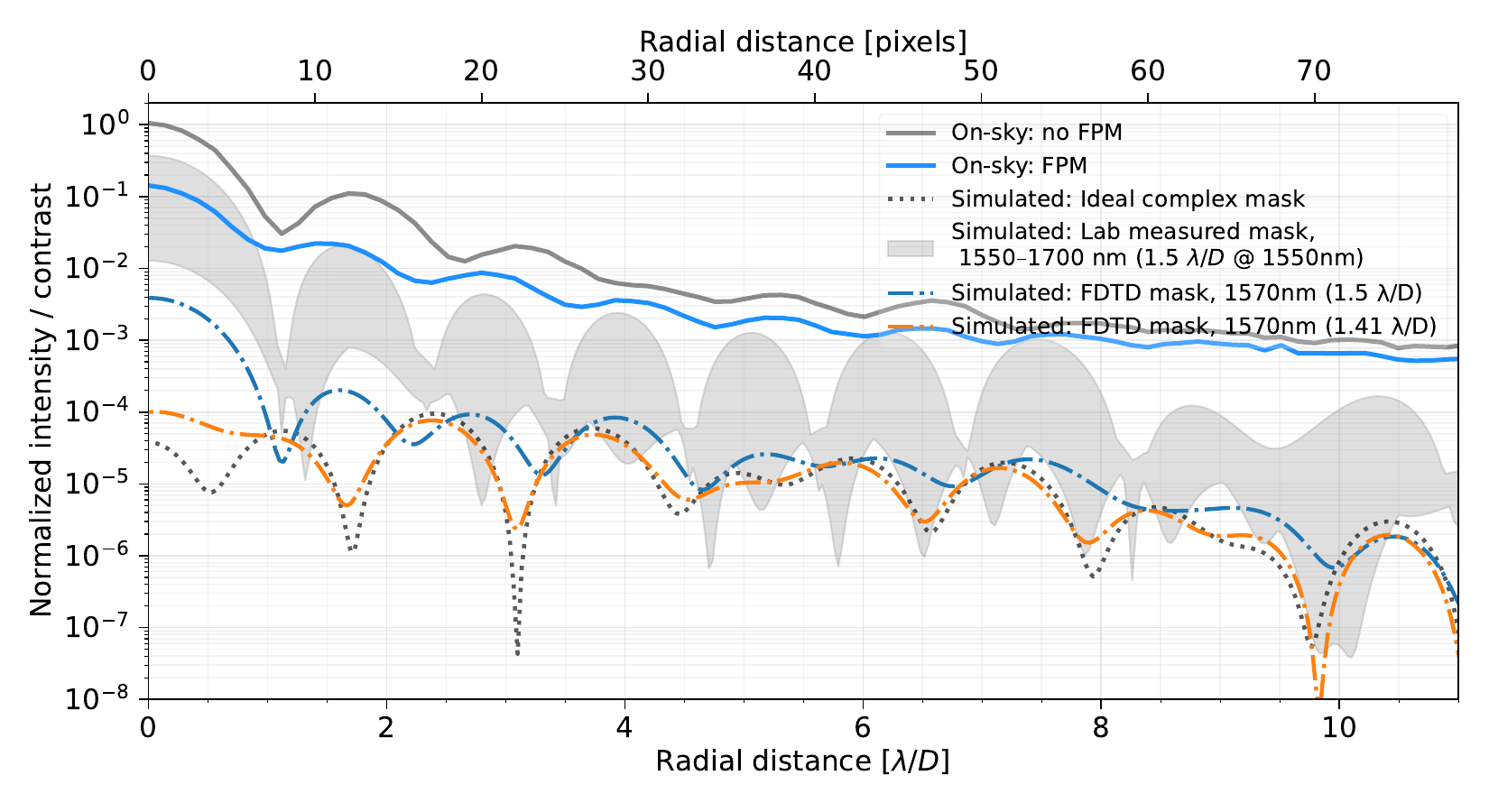}
    \caption{Raw contrast curves derived from the on-sky observations together with simulated performance predictions. The two solid curves show the radial profiles measured on-sky: the black curve corresponds to the non-coronagraphic reference observation (without the focal-plane mask), while the blue curve corresponds to the coronagraphic observation (with the focal-plane mask). The shaded gray region shows the range of predicted performance obtained using the laboratory-measured phase and transmission response of the fabricated metasurface over the wavelength range 1550--1700 nm. The dash-dotted blue and orange curves show simulations using the FDTD-derived phase and transmission response of the metasurface at (1570\,$\mathrm{nm}$) for the manufactured (1.5\,$\lambda/D$) and optimized (1.41\,$\lambda/D$) mask diameters, respectively. The dotted gray curve shows the theoretical performance limit for an ideal complex mask with optimized diameter and perfect complex transmission.}
    \label{fig:radial_profile_comparison}
\end{figure}
The on-sky contrast curves were then calculated by azimuthally averaging the masked images about the primary. To place the on-sky performance in context, several contrast curves were simulated with different FPMs. An end-to-end optical simulation pipeline representative of the MagAO-X instrument was made using HCIPy \cite{por2018hcipy}. It included the MagAO-X pupil geometry, PIAA and inverse-PIAA optics \cite{foster2023implementation} and Lyot stop designed for the instrument, and the hybrid focal-plane mask shown schematically in Figure~\ref{fig:optical_setup}. This would allow the metasurface to be evaluated within the intended PIAA architecture. 

The near-field of the metasurface, obtained from the FDTD simulations, was imported directly as the FPM. Since the mask design and fabrication preceded the development of the end-to-end model, the physical mask diameter of approximately \(160\,\mu\mathrm{m}\) (\(1.5\,\lambda/D\) at \(1550\,\mathrm{nm}\)) was selected a priori rather than through optimization. The simulations were therefore used to assess the expected performance of the manufactured device and provide a direct comparison with the laboratory metrology and on-sky measurements presented in the following sections. Figure~\ref{fig:radial_profile_comparison} compares the predicted performance of four focal-plane mask models with the on-sky performance.

An ideal complex mask with transmission $-1$ and the optimized diameter of $1.41\,\lambda/D$ establishes the theoretical performance limit with the MagAO-X architecture for a simple phase-dot type CMC. Replacing the ideal response with the FDTD-derived response of the fabricated metasurface while retaining the optimized diameter produces nearly identical on-axis performance, with both masks reaching contrasts of approximately $10^{-4}$. This indicates that the metasurface design itself is capable of approaching the ideal solution.

The impact of the fabricated mask diameter was assessed by increasing the focal-plane mask size from the optimal value of $1.41\,\lambda/D$ to the manufactured value of $1.50\,\lambda/D$. This degrades the predicted on-axis contrast from $\sim10^{-4}$ to $\sim4\times10^{-3}$, demonstrating that the device performance is highly sensitive to mask diameter. Substituting the laboratory-measured phase and transmission response further degrades the predicted on-axis contrast to $\sim10^{-2}$, with the best performance occurring near $1650\,\mathrm{nm}$ where the measured complex transmission most closely matches the design target.

The on-sky coronagraphic observations achieve an on-axis contrast of approximately $10^{-1}$ relative to the non-coronagraphic reference image. This performance is broadly consistent with the trends predicted from the laboratory-characterized metasurface response, demonstrating agreement between laboratory metrology, end-to-end simulations, and on-sky measurements. However, the deep destructive-interference minima predicted by the simulations are not observed on-sky owing to residual uncorrected aberrations within the system.

\section{Wavefront-sensing performance in lab}
\label{sec: lab_wfs}

\begin{figure}[hbt!]
    \centering
    \includegraphics[width=\linewidth]{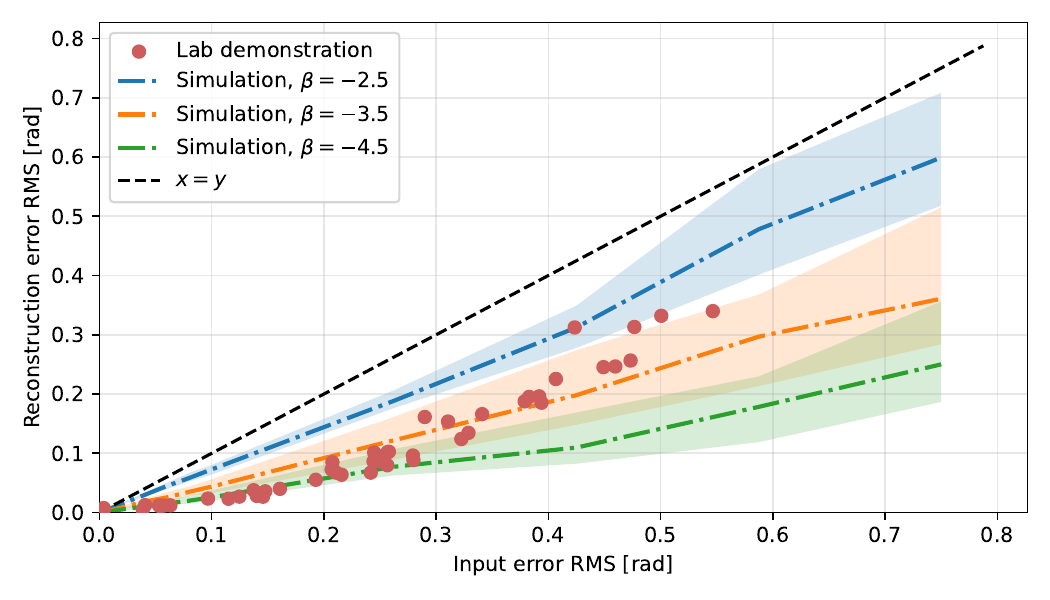}
    \caption{Dynamic range of the metasurface using the MagAO-x's internal source compared to simulations. The laboratory reconstruction error (red markers) is shown for random aberrations generated from linear combinations of the first 10 Zernike modes.
    Simulated reconstruction performance is shown for aberrations generated from power spectral density distributions with exponents ($\beta=-2.5$), $(\beta=-3.5)$, and $(\beta=-4.5)$. For each input aberration RMS, 50 independent phase screens were generated and reconstructed; the curves show the median reconstruction error, while the shaded regions indicate the 16th and 84th percentiles. The dashed black line shows x=y, corresponding to a reconstruction error equal to the input aberration RMS, i.e no improvement.}
    \label{fig:zernike_recon}
\end{figure}

The wavefront-sensing performance of the metasurface was characterized using the MagAO-X internal source at $1300\,\mathrm{nm}$. Aberrations were generated as random linear combinations of Zernike modes, producing 50 realizations spanning a range of input amplitudes. The wavefronts were reconstructed using a linear Zernike reconstructor. Both the injected aberrations and reconstructed coefficients were expressed in deformable mirror (DM) units and converted to radians for comparison with simulations. The conversion assumed a calibration of $1\,\mathrm{\mu m}$ peak-to-valley surface displacement per DM unit, with the corresponding optical path difference converted to phase at $1300\,\mathrm{nm}$. A peak-to-valley to RMS conversion factor of 4.88, representative of the average of the first ten Zernike modes, was used. The resulting reconstruction errors are shown in Figure~\ref{fig:zernike_recon}.

The same exercise was performed in simulation using an ideal ZWFS with a focal-plane mask diameter matching the fabricated metasurface ($\sim1.7\,\lambda/D$ at $1420\,\mathrm{nm}$; see Figure~\ref{fig:fdtd_mask}). For each input RMS, 50 phase screens were generated from a power spectral density (PSD) distribution with $\beta=-2.5$, $-3.5$, and $-4.5$. The wavefronts were reconstructed using the nonlinear reconstructor described in \cite{haffert2024into}, using a basis of the first 10 Zernike modes (excluding piston). The median reconstruction error over 50 independent realizations was then computed.

Figure~\ref{fig:zernike_recon} compares the reconstruction error measured in the laboratory with the simulated results. The laboratory measurements follow a trend similar to the simulations, with noticeable deviations from the low-aberration regime beginning at input aberrations of approximately $0.3\,\mathrm{rad}$. This behavior is consistent with the onset of nonlinearities predicted by the simulations, which occurs between $\sim0.3$ and $0.4\,\mathrm{rad}$ depending on the PSD slope. Overall, the agreement between laboratory measurements and simulations confirms that the fabricated metasurface operates as a Zernike wavefront sensor.

\section{Outlook and future work}
\label{sec: conclusion}
We demonstrated a fast prototyping of a hybrid metasurface that combines coronagraphy and wavefront sensing within a common focal plane. The fabricated device was characterized using laboratory metrology, tested on-sky as a complex-mask coronagraph with MagAO-X, and evaluated as a Zernike wavefront sensor using the instrument's internal source. While the measured on-sky performance was consistent with simulations based on the manufactured device, the physical mask diameter of the designed metasurface was not optimal for the MagAO-X optical architecture. Simulations indicate that a smaller mask size could improve the predicted contrast by nearly two orders of magnitude. In addition, multiple variants with different lithographic exposure doses were fabricated on the same chip, but only one device was characterized in this work. A systematic evaluation of the remaining variants may identify fabrication parameters that further improve performance. Future work will focus on using end-to-end optical simulations and EM simulations to jointly optimize the metasurface's chromatic optical response for optimal coronagraphic and wavefront-sensing performance.

\acknowledgments 
The MagAO-X Phase II upgrade program is made possible
by the generous support of the Heising-Simons Foundation.We are very grateful
for support from the NSF MRI Award \#1625441 (MagAO-X). MagAO-X uses
the CACAO software package, which is supported by NSF Award \#2410616. The authors acknowledge funding from NWO VI.Vidi.233.144 and NWO Award 184.036.004.

\bibliography{report} 
\bibliographystyle{spiebib} 

\end{document}